\documentclass[
reprint,
 amsmath,amssymb,
 aps,
pra
]{revtex4-2}

\usepackage{graphicx}
\usepackage{dcolumn}
\usepackage{bm}
\usepackage{hyperref}
\usepackage{upgreek}

\begin{document}

\preprint{APS/123-QED}

\title{Absence of nematic instability in LiFeAs}

\author{Michael Wissmann$^{1,2}$}
\author{Federico Caglieris$^{1,3,4}$}
\author{Xiaochen Hong$^{1,5}$}
\author{Saicharan Aswartham$^1$}
\author{Anna Vorobyova$^6$}
\author{Igor Morozov$^6$}
\author{Bernd Büchner$^{1,2}$}
\author{Christian Hess$^{1,5,}$}
\email{c.hess@uni-wuppertal.de}
 \affiliation{$^1$Leibniz Institute for Solid State and Materials Research IFW Dresden, 01069 Dresden, Germany}
 \affiliation{$^2$Institute of Solid State Physics, TU Dresden, 01069 Dresden, Germany}
 \affiliation{$^3$University of Genova, via Dodecaneso 33, I-16146 Genova, Italy}
 \affiliation{$^4$CNR-SPIN Genova, via Dodecaneso 33, I-16146 Genova, Italy}
 \affiliation{$^5$Fakultät für Mathematik und Naturwissenschaften, Bergische Universität Wuppertal, 42097 Wuppertal, Germany}
 \affiliation{$^6$Lomonosov Moscow State University, Moscow 119991, Russia}

\date{\today}

\begin{abstract}
The relationship between unconventional superconductivity, antiferromagnetism and nematic order in iron-based superconductors (FeSCs) is still highly debated. In many FeSCs superconductivity is in proximity of a nematically and magnetically ordered state. LiFeAs is an exceptional stoichiometric FeSC becoming superconducting below 18 K, without undergoing a structural or magnetic transition. However, some recent experimental studies suggested the existence of finite nematic fluctuations and even a nematic superconducting state. In this study, we employ elastoresistance as a measure of nematic fluctuations in pristine LiFeAs and compare the findings with the elastoresistance of LiFeAs at low Co and V doping levels as well with that of magnetically and nematically ordering NaFeAs. We find LiFeAs and cobalt-doped LiFeAs far away from a nematic instability.
\end{abstract}

\maketitle

Electronic nematic order is a state that spontaneously breaks the rotational symmetry while preserving translational symmetry.  It has become a subject of increasing attention in the context of the rich phase interplay in iron-based superconductors (FeSCs) \cite{Avci2011, Fernandes2012, Bohmer2016}. In most FeSCs, nematicity occurs in close vicinity of an antiferromagnetically ordered state, which was repeatedly proposed to be intimately connected to the evolution of the unconventional superconductivity \cite{Hirschfeld2011, Fernandes2014, Chubukov2016}. In order to disentangle the relation of superconductivity, magnetism and nematicity, it is important to investigate different FeSCs, especially those which exhibit just one or two of these phases.\\
LiFeAs is one of these unusual representatives of the FeSC-family. The role of nematicity in LiFeAs is to date still strongly debated. It is superconducting in the undoped parent compound with a superconducting transition temperature between 16 K and 18 K \cite{Pitcher2010, Aswartham2011, Putzke2012, Nag2016}. However, several experimental probes including SQUID magnetometry \cite{Chu2009_2, Pitcher2010} and muon-spin rotation \cite{Pratt2009, Wright2013} found no evidence of a magnetic transition or long-range magnetic order. However, it has been reported that LiFeAs could be in principle susceptible to nematicity \cite{Toyoda2018, Sun2019} or even that it undergoes an unusual symmetry-breaking transition \cite{Kushnirenko2020, Yim2018}. In particular, in Ref. \cite{Toyoda2018}, the electric field gradient in NMR measurements was used as an indirect probe of local anisotropies in the crystal structure and a finite albeit very small $\eta$ in LiFeAs has therefore been interpreted as the signature of nematic fluctuations. The presence of electronic-nematic fluctuations has also been suggested from an unusual enhancement of the QPI amplitude in STM measurements \cite{Sun2019}, and even the stabilization of static nematicity in the superconducting state and of smectic electronic order under uniaxial strain have been reported in ARPES \cite{Kushnirenko2020} and STM \cite{Yim2018} measurements.\\
Therefore, the natural role of the nematicity in LiFeAs calls for futher investigation, especially with different probes. One very successful experimental approach to investigate nematic properties in FeSCs is measuring the strain-susceptibility of the resistivity $\eta = \frac{\mathrm{d} }{\mathrm{d} \epsilon} \frac{\Delta \rho}{\rho}$. With this elastoresistance technique, Chu et al. \cite{Chu2012} were able to establish the temperature-dependent behavior of $\eta$ as a proof of the electronic origin of nematicity in FeSCs and a measure of the strength of nematic fluctuations, which has since then been confirmed und used by other research groups \cite{Hosoi2016, Kuo2016, Tanatar2016, Hong2020, Caglieris2021}.
\\
In this work, we investigate nematic fluctuations in LiFeAs using elastoresistance measurements on undoped, electron-doped and hole-doped LiFeAs. In apparent contradiction to the aforementioned reports \cite{Toyoda2018, Sun2019, Kushnirenko2020, Yim2018}, our results clearly indicate that pure LiFeAs is far away from a nematic instability. We compare our results with elastoresistance data on NaFeAs, which is significantly different from LiFeAs.\\
\begin{figure}[h]
\includegraphics[width=0.25\textwidth]{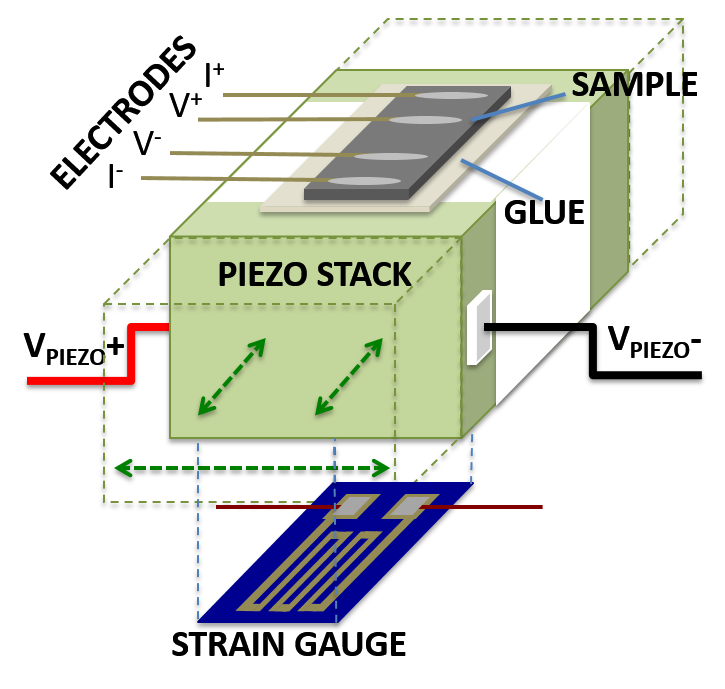}
\caption{\label{setup} Experimental Set-up: Silver wires glued to the sample with silver paint function as electrodes. Green dashed lines at the piezo stack illustrate the expansion under application of a positive voltage $V_\mathrm{p}$. Strain $\epsilon$ is measured using a strain gauge on the backside, figure adapted from Ref. \cite{Hong2020}}
\end{figure}
NaFeAs single crystals were prepared according to Ref. \cite{Steckel2015}. Single crystals of pristine as well as doped LiFeAs were grown by using the self-flux technique as described in Refs. \cite{Morozov2010, Aswartham2011}. Pre-reacted $\mathrm{Li}_3$As, FeAs and $\mathrm{FeAs}_2$ were used. The mixture was placed in a graphite crucible, which was placed in a Nb-container which was later sealed in a quartz ampule to avoid oxidation. This quartz ampule was then placed vertically in a furnace and heated to 1100 $^{\circ}$C. After dwelling for 7 hours, the furnace was cooled down at a rate of 4.5 K/h to 600 $^{\circ}$C and later cooled with 100 $^{\circ}$C/h to room temperature. As-grown crystals (exemplary image of Li($\mathrm{Fe}_{0.972}\mathrm{V}_{0.028})$As in the inset of \autoref{fig1} (d)) are highly sensitive to air and moisture. Single crystals were characterized by using SEM/EDX and powder x-ray diffraction, which confirm the stoichiometry and the phase of the LiFeAs.\\
All sample preparation steps have been performed in an argon-filled glove box with $\mathrm{O}_2$ and $\mathrm{H}_2\mathrm{O}$ content less than 0.3 ppm. The glove box is equipped with a second 2m-long chamber that enables the introduction of a whole measurement probe into the Ar-atmosphere. Few selected pieces of as grown single crystals were cut mechanically into rectangular shapes with lengths $<$ 4 mm and cleaved to thicknesses of the order of 50 $\upmu$m, to ensure a homogeneous transmission of strain through the whole thickness (compare Refs. \cite{Chu2012, Hong2020}). Ag-wires were glued on top of a sample, in order to perform four-terminal resistance measurements. The samples were glued (using Devcon 14250 5 Minute Epoxy) along the [110]-crystal axis on top of a piezoelectric actuator that allowed precise strain-control depending on an applied voltage $V_\mathrm{p}$. The strain $\epsilon$ was measured using a strain gauge glued onto the backside of the piezo (\autoref{setup}). After making contacts to the sample, it was mounted directly onto the probe, which then was closed inside the Ar-box, discharged through the air lock and immediately connected to a high-vacuum pump, which creates a vacuum of less than 1 $\cdot$ $10^{-6}$ mbar to 1 $\cdot$ $10^{-7}$ mbar. The evacuated probe was inserted into liquid helium and cooled-down. Due to cryo-pumping, the sample stayed in cryogenic vacuum conditions until it got warmed-up again.\\
In \autoref{fig1} (a) and (b), we present the temperature-dependent resistivities $\rho\left(T\right)$ of LiFeAs and NaFeAs, respectively.

\begin{figure}[htbp]
   \centering
   \includegraphics[width=0.45\textwidth]{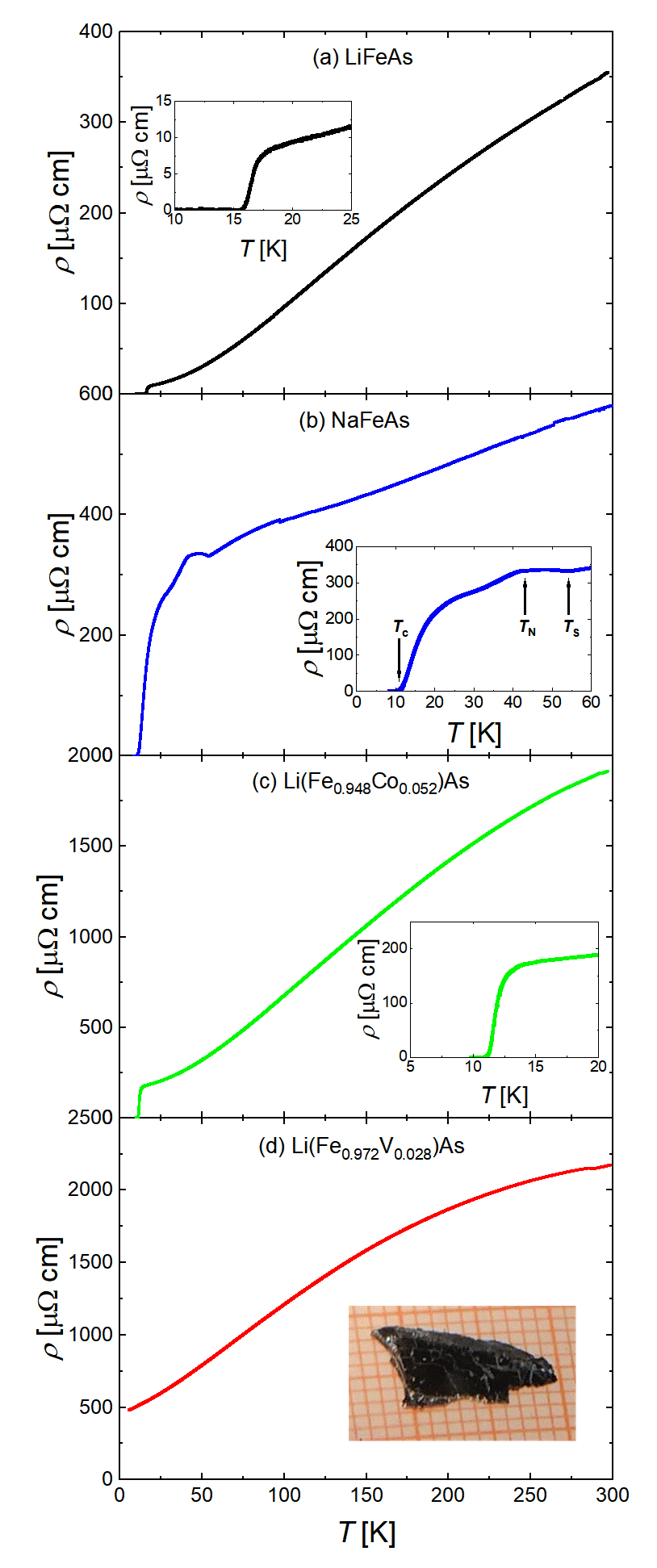}
        \caption{Temperature-dependence of the resistivities $\rho(T)$ of \textbf{(a)} LiFeAs, \textbf{(b)} NaFeAs, \textbf{(c)} Li($\mathrm{Fe}_{0.948}\mathrm{Co}_{0.052})$As, \textbf{(d)} Li($\mathrm{Fe}_{0.972}\mathrm{V}_{0.028})$As; Insets of \textbf{(a), (b), (c)}: Superconducting transitions, Inset of \textbf{(d)}: As-grown single crystal of Li($\mathrm{Fe}_{0.972}\mathrm{V}_{0.028})$As}
        \label{fig1}
\end{figure}

\begin{figure*}[ht]
   \centering
      \includegraphics[width=0.95\textwidth]{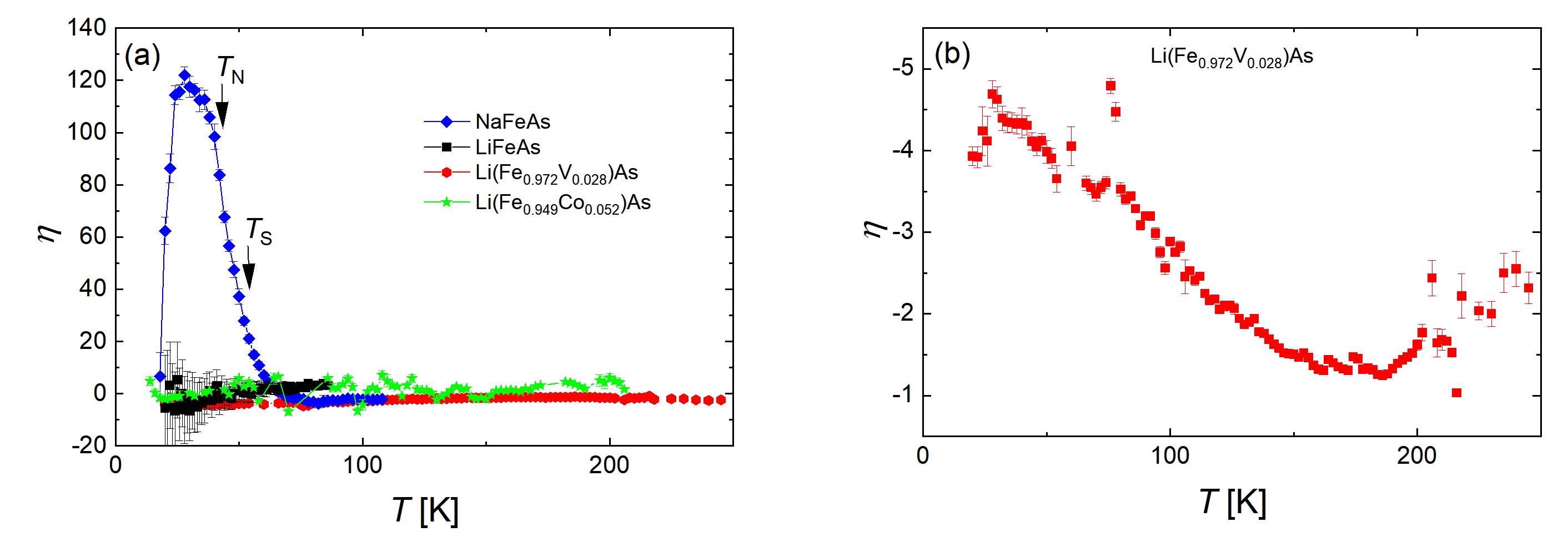}
\caption{\label{fig2} \textbf{(a)} Elastoresistance $\eta$ of NaFeAs and Li($\mathrm{Fe}_{1-x}\mathrm{TM}_{x})$As (TM = V, Co). NaFeAs shows a strong divergence, indicating the increasing amount of nematic fluctuations towards the structural transition temperature. The electronic response to strain is significantly smaller, almost non-existent in Li($\mathrm{Fe}_{1-x}\mathrm{TM}_{x})$As (TM = V, Co). \textbf{(b)} The amplitude of the elastoresistance of Li($\mathrm{Fe}_{0.972}\mathrm{V}_{0.028})$As shows a small increase towards lower temperatures. The error bars of each data point are derived from the statistical error of a linear fit through the  $\frac{\Delta \rho}{\rho}$ vs. $\epsilon$ data. In the tetragonal phase, a negative sign of the nematic susceptibility indicates that the resistance is smaller along the longer crystal axis a than along the shorter b}
\end{figure*}
The resistivity trend and absolute value of both compounds correspond well to earlier reports \cite{Zhang2012,Rullier-Albenque2012}.
The structural and magnetic phase transitions of NaFeAs at $T_\mathrm{S} \sim$ 54 K and $T_\mathrm{N} \sim$ 43 K are visible as kinks in $\rho(T)$ (inset in \autoref{fig1} (b)), whereas no kinks or anomalies are observed in the resistivity of LiFeAs as it approaches the sharp superconducting transition at $T_\mathrm{c} \sim$ 16 K (inset in \autoref{fig1} (a)). This comparably high transition temperature and the small absolute value of $\rho$ indicate a high quality of the sample.\\
In LiFeAs, 5.2 \% Co-doping increases the resistivity and lowers the superconducting transition temperature to $T_\mathrm{c} \sim$ 11 K (\autoref{fig1} (c)).  Apparently, as can be inferred from \autoref{fig1} (d), the impact of V-doping on the resistivity is much stronger than that of Co-doping. Already at 2.8 \% V-doping the resistivity is higher than that of the Co-doped sample and no sign of a superconducting transition down to 5 K is observed. The $\rho-T-$curves of both doped LiFeAs compounds show no indication of a magnetic or nematic phase transition, consistent with previous reports \cite{Xing2014, Xing2016, Xu2020sup}.\\
In \autoref{fig2} (a), the temperature dependence of the elastoresistance $\eta$ of all four FeSCs investigated in this study is shown.
Remarkably, only the elastoresistance amplitude of NaFeAs shows a diverging behavior towards low temperatures typical for FeSCs with a nematic phase transition, indicating the increase of nematic fluctuations towards $T_\mathrm{S}$. In contrast, in LiFeAs, at no temperature a clear electronic response $\frac{\Delta \rho}{\rho}$ to strain is observed. This is direct evidence for the absence of nematic fluctuations, since local structural distortions connected to local nematic order evidentially change the strain response of the resistivity. The lack of a measurable increase of the elastoresistance amplitude towards low temperatures indicates that the material is far away from a nematic instability.\\
Having established that pure LiFeAs is far away from a nematic instability opens up the question, whether it is possible to drive the material towards nematicity by tuning external parameters. Here, we investigate the effect of electron- and hole-doping on the nematic properties of LiFeAs by measuring elastoresistance of Li($\mathrm{Fe}_{0.948}\mathrm{Co}_{0.052})$As and Li($\mathrm{Fe}_{0.972}\mathrm{V}_{0.028})$As, respectively.
In 5.2 \% Co-doped LiFeAs, similar to the undoped parent compound, no electronic response to strain could be measured which leads to elastoresistance values scattered around zero. The absence of any sign of local structural distortion/nematic fluctuation in LiFeAs does not change upon 5.2 \% Co-doping.\\
On the other hand, we report a small divergence of the nematic susceptibility towards low temperatures in Li($\mathrm{Fe}_{0.972}\mathrm{V}_{0.028})$As (\autoref{fig2} (b)). Even though the absolute amplitude is low, indicating a small but finite amount of nematic fluctuations, a trend is clearly visible. Note, that the low temperature absolute value of $\rho$ of Li($\mathrm{Fe}_{0.972}\mathrm{V}_{0.028})$As is an order of magnitude higher compared to LiFeAs, thus enabling a very precise measurement of $\eta$ - as also noticable in the comparatively small size of the error bars.\\ 
The absence of nematicity in elastoresistance measurements on LiFeAs might seem at odds with the aforementioned NMR \cite{Toyoda2018}, ARPES \cite{Kushnirenko2020} and STM \cite{Sun2019, Yim2018} reports. However, a closer inspection reveals a good consistency with these data. More specifically, in Ref. \cite{Toyoda2018}, the electric field gradient of LiFeAs is reported to be more than one order of magnitude smaller than that of NaFeAs (absolute value smaller than 0.005 even at low temperatures). Such a small-amplitude sign of electronic anisotropy is in principle consistent with the small increase of the elastoresistance signal in LiFeAs in \autoref{fig2}. However, also given the significant range of error compared to the small absolute values, the signal can not unambigously be assigned to nematicity.
Furthermore, the nematic order reported from ARPES in Ref. \cite{Kushnirenko2020} is interpreted as a result of the superconductivity and therefore should disappear at the elastoresistance-relevant temperatures above $T_\mathrm{c}$ (within the limits of superconducting fluctuations).
Finally, the QPI measurements and analysis in Ref. \cite{Sun2019} suggest a coupling between electronic and supposedly nematic lattice modes at energies in the range of 10 meV. The energy scale of our elastoresistance measurements is expected to be smaller since we measure in quasistatic conditions in the zero-strain limit. It seems reasonable to expect that under these condictions the nematic modes are energetically too far away to affect the transport data.\\
Therefore in consistency with these previous reports on nematic properties in LiFeAs, we conclude that 
the missing strain-response in the elastoresistance measurements presented above strongly suggests that pure, undoped LiFeAs is far away from a nematic instability.\\
The results on Co-doped LiFeAs seem in principle consistent with the observation on canonical FeSCs with nematic ordering parent compounds (i.a. Ba$\mathrm{Fe}_2\mathrm{As}_2$, NaFeAs) that the amplitude of the nematic fluctuations declines upon increasing Co-doping to the overdoped side (as seen in elastotransport \cite{Kuo2016} and NMR \cite{Toyoda2018}). In such a scenario, LiFeAs would be located deep in the electron doped side of the electronic phase diagram. In this case, hole doping should in principle drive the system closer to the nematic phase with a higher amplitude of nematic fluctuations.
Indeed, this is what is observed upon V-doping, where vanadium could act as an effective hole dopant. In fact, V-doping is known to promote the emergence of strong antiferromagnetic spin fluctuations, which seem supportive of this scenario \cite{Xing2016, Xu2020, Sheng2020}. However, at the same time the V-doping apparently suppresses the superconductivity, which indicates that this doping scheme does not simply drive the system towards lower doping levels in the phase diagram of the canonical electron-doped FeSCs.\\
In summary, we employed elastoresistance as direct probe of the nematic properties of undoped, electron-doped and hole-doped LiFeAs. We showed, that there is no sign of nematic fluctuations detectable in LiFeAs. Furthermore, Co-doping does not induce nematic fluctuations, indicating that LiFeAs is on the electron-doped side far away from a nematic instability. Finally, V-doping of LiFeAs might induce nematic fluctuations, while at the same time suppressing superconductivity.\\
This project has been supported by the Deutsche Forschungsgemeinschaft through the Research Projects CA 1931/1-1 (F.C.) and AS 523/4-1 (S.A.). Furthermore, this project has received funding from the European Research Council (ERC) under the European Union’s Horizon 2020 research and innovation programme (Grant Agreement No. 647276-MARS-ERC-2014-CoG) (X.H., C.H.). I.M., A.V., S.A. and B.B. thank DFG and RSF for financial support in the frame of the joint DFG-RSF project “Weyl and Dirac semimetals and beyond – prediction, synthesis and characterization of new semimetals".

\appendix

\bibliography{collection}

\end{document}